\documentclass[twocolumn,10pt,superscriptaddress]{revtex4-2}
\usepackage[plainpages=false,pdfpagelabels,colorlinks=true,linkcolor=red,urlcolor=blue,citecolor=blue,pdftitle={Title},pdfauthor={},pdfdisplaydoctitle=true,pdfduplex=DuplexFlipLongEdge]{hyperref}
\usepackage{amsmath, amsthm, amssymb}
\usepackage{array}
\usepackage{bbold} 
\usepackage{bm}        
\usepackage[dvipsnames,usenames]{color}
\usepackage{dcolumn}
\usepackage{float}
\usepackage{graphicx}  
\usepackage{lipsum}
\usepackage{mathrsfs}
\usepackage{mathtools,amssymb,graphicx,units,bm}
\usepackage{physics}%
\usepackage{soul} 
\usepackage{booktabs}
\usepackage[normalem]{ulem}
\usepackage{subcaption}

\graphicspath{{figures/}}

\emergencystretch=\maxdimen
\usepackage[utf8]{inputenc}
\DeclareUnicodeCharacter{202F}{\,} 

\begin{document}

\normalem

\title{\large \textbf{A Swampland-modified Hod bound for charged black holes with exotic matter}}

\author{S. Saoud}
    \email{$soulaimane_saoud@um5.ac.ma$}
    \affiliation{LPHE-MS, Science Faculty, Mohammed V University in Rabat, Morocco}
    \affiliation{Centre of Physics and Mathematics, CPM-Morocco}
    
\author{M. A. Rbah}
    \email{$Mohamedamin_rbah@um5.ac.ma$}
    \affiliation{LPHE-MS, Science Faculty , Mohammed V University in Rabat, Morocco}
    \affiliation{Centre of Physics and Mathematics, CPM-Morocco}
    
\author{R. Sammani}
    \email{$rajae_sammani@um5.ac.ma$}
    \affiliation{LPHE-MS, Science Faculty , Mohammed V University in Rabat, Morocco}
    \affiliation{Centre of Physics and Mathematics, CPM-Morocco}
    
\author{E. H. Saidi}
    \email{$ e.saidi@um5r.ac.ma$}
    \affiliation{LPHE-MS, Science Faculty , Mohammed V University in Rabat, Morocco}
    \affiliation{Hassan II Academy of Science and Technology, Kingdom of Morocco}
    \affiliation{Centre of Physics and Mathematics, CPM-Morocco}
    
\author{R. Ahl Laamara}
    \email{$r.ahllaamara@um5r.ac.ma$}
    \affiliation{LPHE-MS, Science Faculty , Mohammed V University in Rabat, Morocco}
    \affiliation{Centre of Physics and Mathematics, CPM-Morocco}
\date{\today}

\begin{abstract}
In this paper, we study the quasinormal modes (QNMs) of a charged black hole in the presence of both quintessence and a cloud of strings by using the Padé-averaged higher-order WKB approximation method. We investigate the effect of the quintessence parameter $\alpha$ and the cloud of strings parameter $\lambda$ on stability as well as the oscillation frequency of perturbations. The validity of Hod’s conjecture, which relates quasinormal frequencies to the black hole temperature, is tested throughout the physically allowed parameter space. Our results show that both the effective potential and the decay rate of perturbations depend on the value of $\alpha$ and $\lambda,$ leading to either enhancement or suppression of the conditions required to satisfy Hod’s bound. Furthermore, we discuss how these parameters modify the black hole shadow and the corresponding energy emission rate, proving correlations with observable signatures. Finally, we establish the connection with the Swampland Distance Conjecture by expressing the Hawking temperature in terms of the scalar field excursion. Our analysis leads to a modified Hod bound and identifies a region of parameter space in which both the modified Hod bound and the Swampland constraints are simultaneously satisfied, ensuring consistency between black hole thermodynamics, observational properties, and quantum gravity constraints.
\end{abstract}
\maketitle

\section{Introduction}
Black holes are among the most profound predictions of Einstein’s general relativity, providing a natural theater in which the interaction between gravity, quantum theory, and thermodynamics can be explored. Understanding their perturbative and thermodynamic properties has been a crucial facet in the exploration of the microscopic structure of spacetime and remains a cornerstone in ongoing attempts to probe quantum aspects of gravity. In particular, the behavior of quasinormal modes (QNMs) encodes information about the dynamical response and relaxation process of black holes \cite{Konoplya:2011qq, Konoplya:2019hlu,Iyer:1986np,Matyjasek:2019eeu} while their optical features like shadows and emission spectra have become powerful probes of near-horizon geometry
\cite{Wei_2013,Cuadros-Melgar:2020kqn, Gogoi:2024vcx,EventHorizonTelescope:2019dse,EventHorizonTelescope:2022wkp,Belmahi:2024sqp}.\\
In recent years, a considerable effort has been devoted to studying black holes immersed in non-trivial matter distributions such as quintessence fields and stringy clouds, which modify both the asymptotic structure of spacetime and its near-horizon thermodynamics. Quintessence, described by a dynamical scalar field with the pressure equation of state $p_q = \omega_q \rho_q$, has been extensively investigated as a candidate for cosmological dark energy\cite{Kiselev:2002dx, Tsujikawa:2013fta, Caldwell:1997ii, Copeland:1997et, Ferreira:1997hj}. Its appearance near the event horizon of black holes leads to new horizon structures and significantly modifies their evaporation properties \cite{Toledo:2018hav, Saleh_2011, Liu:2025dhl, Al-Badawi:2024mco}. On the other hand, a surrounding cloud of strings introduces a homogeneous energy density and a non-trivial topology that deforms the metric function and affects both the stability and optical appearance of the black hole \cite{PhysRevD.20.1294, Zahid:2023csk, Sun:2022wya, Raza:2025ohk}. Accordingly, the simultaneous presence of both of these exotic ingredients provides a promising framework to investigate how additional fields and matter sectors influence black hole dynamics and X-ray astrophysical signatures, beyond their impact on thermodynamics and relaxation behavior alone.
More fundamentally, quantum gravity considerations suggest that not all EFTs coupled to gravity can be uplifted to high energies. The \textit{Swampland Program} aims to separate out the set of low-energy EFTs which have a consistent embedding into a quantum theory of gravity (known as the ``Landscape’’) from those that do not (the “Swampland”) \cite{Ooguri:2006in, Palti:2019pca, Sammani:2025hpz, Sammani:2025ydx,Charkaoui:2024uzi}. One of its central conjectures, the \textit{Swampland Distance Conjecture} (SDC) \cite{Ooguri:2006in,Ooguri:2018wrx,Kehagias2019ANO,Brahma:2019kch,Cicoli:2021skd,Storm:2020gtv}, states that when a scalar field explores trans-Planckian distances in moduli space, an infinite tower of exponentially light states emerges signaling the breakdown of the EFT description. This conjecture, which was originally formulated in the context of string theory, has far--reaching consequences for cosmology and black hole physics, particularly for models in which scalar fields (such as quintessence) can evolve over very large field-ranges.\\
From a complementary standpoint, the \textit{Hod conjecture} \cite{Hod:2006jw} imposes a universal upper bound on the decay rate of black hole quasinormal modes,
$ 
|\mathrm{Im}(\omega)| \leq \pi T_H ,
$ The dynamical dissipation of black hole perturbations provides a general relationship between the decay timescale and the Hawking temperature $T_H$. This constraint describes an upper limit on the rate of dissipation of black hole perturbations, given that there exists a consistent thermodynamic framework for black holes
 \cite{Konoplya_2022, Bonanno:2025dry, Sekhmani:2023ict, Lambiase:2023hng, Parbin:2022iwt}. Interestingly, recent studies have uncovered intriguing connections between Swampland criteria and QNM properties, suggesting a possible deep consistency relation between semiclassical black hole physics and quantum gravity constraints\cite{Andriot:2025los, Sammani:2025aat}.\\
In this work, we study the interplay between Hod and SDC in the context of a charged Reissner–Nordström black hole simultaneously surrounded by a quintessence field and string cloud. Exploiting the metric structure developed in \cite{Kiselev:2002dx, PhysRevD.20.1294}, we study how the quintessence parameters $\alpha$ and the cloud of strings intensity $\lambda$ modify the effective potential, quasinormal spectrum, and thermodynamic properties of the black hole. Using the sixth-order Padé-averaged WKB method \cite{Matyjasek:2019eeu, Konoplya:2019hlu, Konoplya:2003ii}, we compute numerical values of the QNM modes and derive a generalized Hod bound in terms of a field excursion $\Delta\phi$ imposed by the SDC. Moreover, by investigating the overlap of the parameter space allowed by Hod's and the  SDC inequalities, we identify physically viable regions in which both thermodynamic consistency and quantum-gravity requirement can be concurrently satisfied. \\
The work is organized as follows. In Section~\ref{sec:model}, we present the Reissner--Nordström black hole solution surrounded by both quintessence and a cloud of strings, and highlight the key features of the spacetime geometry. In Section~\ref{sec:develop}, we analyze the scalar perturbations and compute the corresponding quasinormal modes, to examine the validity of Hod’s conjecture in this background. The optical properties of the black hole, including its shadow and emission rate, are explored in Section~\ref{sec:opp}. In Section~\ref{sec:hsdb}, we establish the connection between Hod’s bound and the Swampland Distance Conjecture, and discuss the quantum gravity implications of our results. Finally, Section~\ref{sec:conclusions} summarizes our main findings and highlights possible future extensions of this work.

\section{Reissner-Nordström black holes with quintessence and a cloud of strings} \label{sec:model}
The Reissner–Nordström (RN) black hole is the simplest static, spherically symmetric solution to Einstein’s field equations describing a charged, non-rotating black hole in asymptotically flat spacetime. Its geometry is uniquely determined by the balance between gravitational and electromagnetic interactions, and the radius in Schwarzschild-like coordinates depends solely on two parameters, the BH mass \( M \) and its electric charge \( Q \).

To generalize such a classical setup, we introduce two more exotic matter sources, a quintessence field and a cloud of strings. These sources do not contribute trivially to the total stress–energy tensor, and so modify the Einstein equations and cause extra terms in the metric function. The obtained geometry is a generalization of Reissner–Nordström spacetime, which describes the combined effect of gravity due to charge, quintessence and string cloud matter.

In this generalized framework, the metric is given by
\begin{equation}
ds^2 = -f(r) \, dt^2 + \frac{dr^2}{f(r)} + r^2 \left( d\theta^2 + \sin^2\theta \, d\phi^2 \right),
\end{equation}

where the metric function \( f(r) \) takes the form

\begin{equation} \label{eq:f(r)}
f(r) = 1 - \frac{2M}{r} + \frac{Q^2}{r^2} - \lambda - \frac{\alpha}{r^{3\omega_q + 1}}.
\end{equation}
The parameter \(\lambda\) is related to the cloud of strings, describing a homogeneous energy density that distorts the asymptotic structure of spacetime~\cite{PhysRevD.20.1294,Toledo:2018hav}. Physically, the string cloud can be thought of as an array of one-dimensional objects distributed along spacetime, which give rise to a uniform contribution to the stress-energy tensor and therefore modify the horizon structure.
The parameter \( \alpha \) characterizes the intensity of the quintessence field, with the equation of state given by

\begin{equation}
p_q = \omega_q \rho_q,
\end{equation}

with $-1 \leq \omega_q \leq -\tfrac{1}{3}$~\cite{Kiselev:2002dx,Saleh_2011}. The quintessence component behaves as a cosmic fluid and introduces a power-law correction term into the large-scale (asymptotic) geometry of spacetime. This term can significantly affect the asymptotic structure and may even lead to the emergence of additional horizons, depending on the value of \( \omega_q \).
The composite stress–energy tensor of the quintessence, string cloud, and electromagnetic fields satisfies Einstein’s equations, resulting in a spacetime geometry richer than that of the standard RN BH. These additional contributions, in the extended metric, not only deform the background but also affect the propagation of perturbations, gravitational, electromagnetic, and scalar. On this extended spacetime, the horizons are determined by the roots of \( f(r) = 0 \), which is generally more intricate than in the Schwarzschild or Reissner--Nordström cases due to the presence of the extra parameters \( \lambda \), \( \alpha \), and \( \omega_q \).
This model, therefore, provides a robust framework to study thermodynamic properties, stability, and dynamical issues of black holes in the presence of exotic matter fields. In particular, the effects of parameters \( \lambda \), \( \alpha \), and \( \omega_q \) on the QNM spectra and optical properties will be studied in detail subsequently~\cite{Toledo:2018hav,Saleh_2011}.

\section{Quasinormal Modes and Hod's Conjecture} \label{sec:develop}
\label{sec:qnm}

In this section, we study scalar perturbations of a Reissner--Nordström black hole bathed in a cloud of strings and quintessence. We are interested in finding the associated quasinormal mode (QNM) spectrum, using the higher-order Padé averaged WKB approximation and examining the validity of Hod's conjecture in this particular framework.

\subsection{Dynamics of scalar fields}
The dynamics of a massless scalar field denoted as $\Psi$ in a curved background is determined by the Klein--Gordon equation

\begin{equation}\label{eq4}
\square \Psi = \frac{1}{\sqrt{-g}} \partial_\mu \left( \sqrt{-g}\, g^{\mu\nu} \partial_\nu \Psi \right) = 0,
\end{equation}
where $\square$ denotes the d’Alembert operator. 

To separate the temporal, radial, and angular dependencies, we adopt the standard ansatz

\begin{equation}
\Psi(t,r,\theta,\phi) = e^{-i\omega t} R(r) Y_{\ell m}(\theta,\phi),
\end{equation}
where $\omega$ is the oscillation frequency, $Y_{\ell m}(\theta,\phi)$ are the spherical harmonics, with $\ell$ and $m$ denoting the angular momentum quantum numbers. Substituting this ansatz into Eq.~\eqref{eq4} yields the radial equation  

\begin{equation}
\frac{d^2R}{dr^2}
+ \left( \frac{2}{r} + \frac{f'(r)}{f(r)} \right)\frac{dR}{dr}
+ \left( \frac{\omega^2}{f(r)^2} - \frac{\ell(\ell+1)}{r^2 f(r)} \right) R = 0,
\end{equation}

where $f(r)$ is the metric function given in Eq.~\eqref{eq:f(r)}.

\subsection{Wave Equation in Schrödinger-like Form}

Introducing the tortoise coordinate $dr_* = \frac{dr}{f(r)}$ and redefining the radial function as $R(r) = \psi(r)/r$, the radial equation can be recast into a Schrödinger-like wave equation  

\begin{equation}
\frac{d^2\psi}{dr_*^2} + \left[ \omega^2 - V(r) \right] \psi = 0,
\label{eq:wave_eq}
\end{equation}
where the effective potential is expressed as  

\begin{equation}
V(r) = f(r)\left( \frac{\ell(\ell+1)}{r^2} + \frac{f'(r)}{r} \right).
\label{eq:potential}
\end{equation}

\subsection{Qualitative behavior of the Effective Potential}
The azimuthal momentum number \( \ell \) controls the strength of the centrifugal barrier in the effective potential. Increasing \( \ell \) significantly raises the height of the potential barrier and slightly shifts its maximum closer to the black hole horizon. Such reliance on \( \ell \) is a common feature of spherically symmetric black hole backgrounds, and has a direct impact on the spectrum of QNMs frequencies.
\begin{figure}[H]
   \centering
   \begin{subfigure}{0.35\textwidth}
       \includegraphics[width=\linewidth]{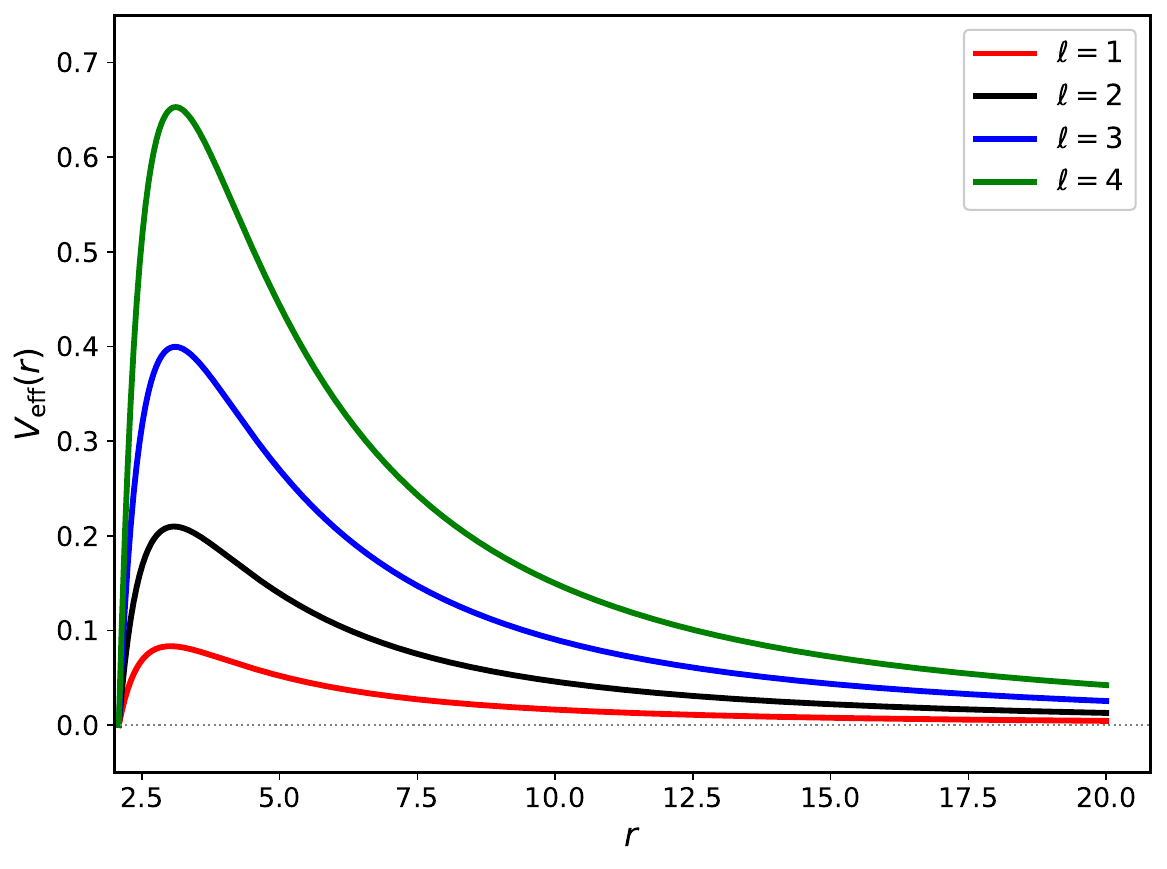}
       \caption{ $\lambda=0.05$, $\alpha=0.01$}
       \label{figch3033:SHAD0.01.png}
   \end{subfigure}
   \hfill
   \begin{subfigure}{0.35\textwidth}
       \includegraphics[width=\linewidth]{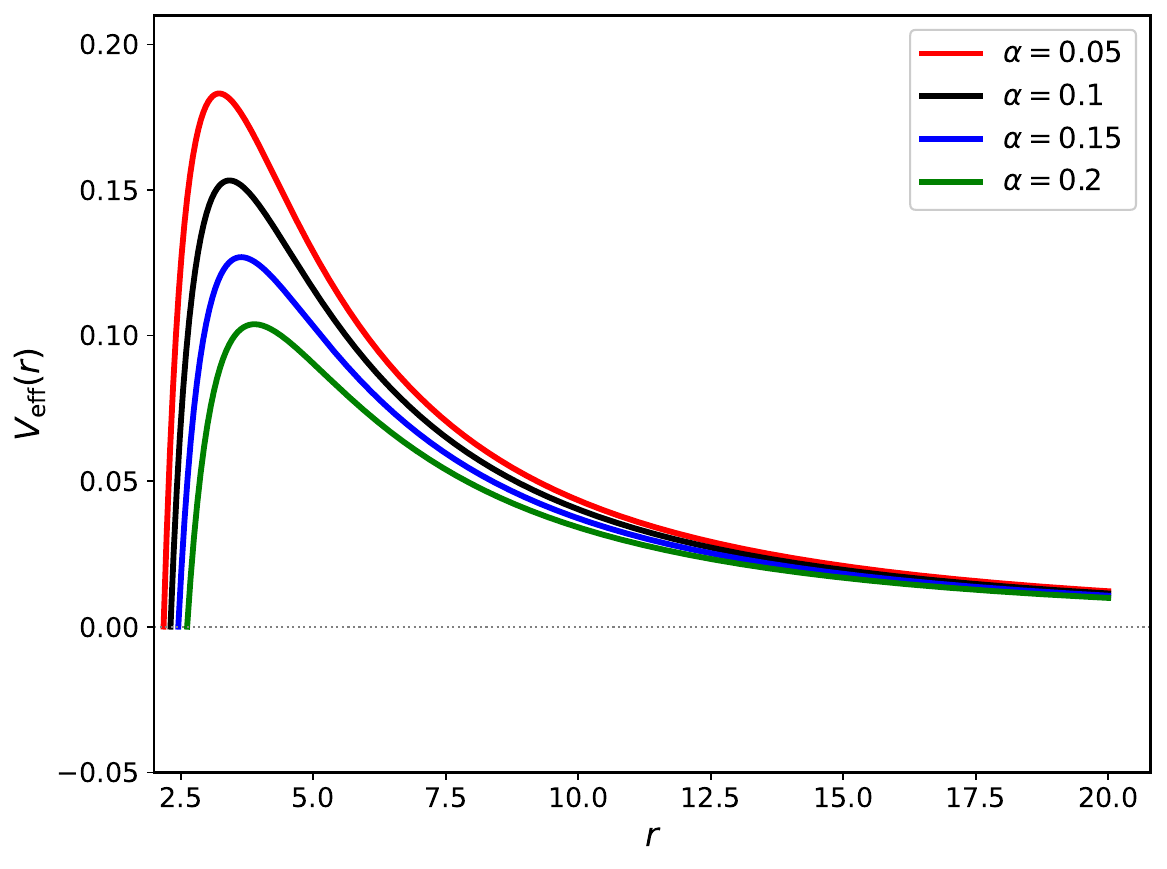}
       \caption{ $\lambda=0.05$, $\ell=2$ }
       \label{figch3033:SHd0.08.png}
   \end{subfigure}
   \hfill
   \begin{subfigure}{0.35\textwidth}
       \includegraphics[width=\linewidth]{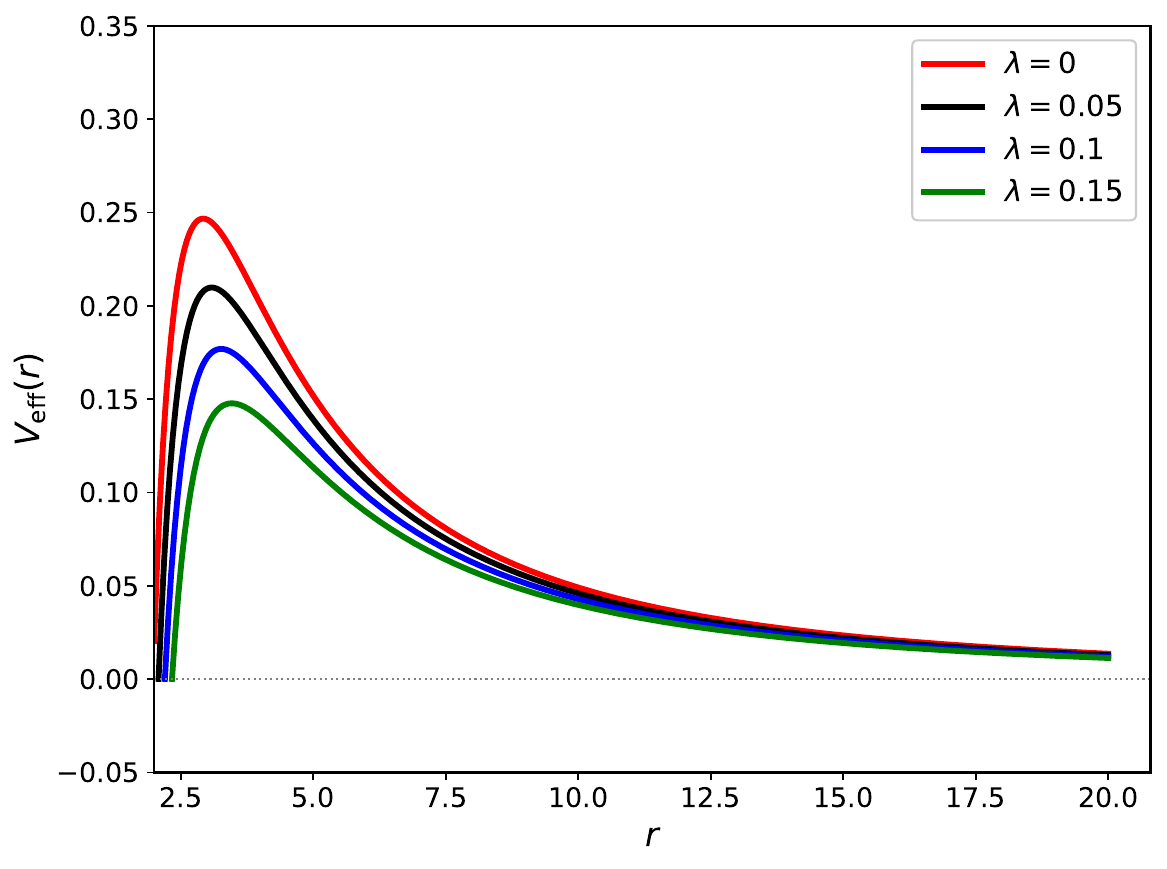}
       \caption{  $\alpha=0.01$, $\ell=2$}
       \label{figch3033:SHd8.png}
   \end{subfigure}
   \caption{The effective potential for scalar perturbations in RN black holes with quintessence and a cloud of strings is shown for the parameter choices $M=1$, $Q=0.3$, and $\omega_q=-\tfrac{1}{3}$.}
   \label{fig:matrice_figuresss}
\end{figure} 

As illustrated in Figs. (a) and (b), variations in the quintessence parameter $\alpha$ and the string cloud variable $\lambda$ substantially modify the effective potential $V_{\text{eff}}(r)$. Increasing either one lowers the height of the barrier and broadens its profile, while the peak shifts slightly outward. These effects reflect the repulsive role played by quintessence to perturbations, as well as the weakening of gravity associated with the cloud of strings that diminishes confinement effects around the BH. Therefore, the scalar perturbations decay more rapidly, suppressing long-lived oscillatory modes and shortening the relaxation time of the system.
\subsection{Padé averaged WKB approximation method}
To compute the QNM frequencies, we employ the sixth-order Padé averaged Wentzel-Kramers-Brillouin (WKB) method. The method has been used to obtain reliable values of oscillation frequencies of GWs in different astrophysical scenarios \cite{Konoplya:2019hlu,Konoplya:2011qq, Konoplya:2003ii,Matyjasek:2019eeu,Konoplya_2022}. The sixth-order WKB method has refined the numerical results to better accommodate observations.
The complex frequency \(\omega\) can be accordingly given by

\begin{equation}\label{9}
    \omega = \sqrt{-i \left[ (n + 1/2) + \sum_{k=2}^{6} \bar{\lambda}_{k} \right] \sqrt{-2 V_{0}^{\prime\prime}} } + V_{0},
\end{equation}
where $n = 0, 1, 2, \ldots$ corresponds to the modes (or harmonics), and \(V_{0}\) refers to the value of the potential function \(V_{s}\) at its maximum radius.  \(V_{0}^{\prime\prime}\) is its second derivative also evaluated at \(r = r_{\text{max}}\). 
The correction terms \(\bar{\lambda}_{k}\) encode higher-order effects like mode mixing, which are necessary for a precise determination of the oscillation frequencies of GWs in various astrophysical environments \cite{Konoplya:2003ii,Matyjasek:2019eeu,Autthisin:2025ugf,Iyer:1986np}.

\begin{table}[!ht]
   \centering
\begin{tabular}{|c|c|c|c|}
\hline
$\ell$ & Padé Averaged WKB $\omega$ & $\Delta_{\text{rms}}$ & $\Delta_6$ \\
\hline
1 & $0.332883 - 0.056489i$ & 0.0182950917 & 0.0129365834 \\
2 & $0.531795 - 0.066105i$ & 0.0111182631 & 0.00786179921 \\
3 & $0.734706 - 0.071275i$ & 0.00736369505 & 0.00520691870 \\
4 & $0.939049 - 0.073370i$ & 0.00584832322 & 0.00413538901 \\
5 & $1.144152 - 0.074193i$ & 0.00525061990 & 0.00371274894 \\
6 & $1.349773 - 0.075688i$ & 0.00418464453 & 0.00295899053 \\
\hline
\end{tabular}
\caption{Quasinormal modes of the black hole with $n = 0$, $M = 1$, $\alpha = 0.1$ and $\lambda = 0.05$.}
    \label{tab:WKB}
\end{table}
Table \ref{tab:WKB} lists the fundamental (n=0) QNMs for various values of $\ell$.

The root-mean-square error $\Delta_{\mathrm{rms}}$ and the quantity $\Delta_6$, defined as
$
\Delta_6 = \frac{\omega_7 - \omega_5}{2}
$,
quantify the convergence of the WKB expansion. As expected, the accuracy improves for higher multipole numbers $\ell$, which is a well-known feature of WKB-type methods. Nevertheless, this approximation remains reliable only when $n < \ell$. \cite{Bonanno:2025dry,Sekhmani:2023ict,Lambiase:2023hng,Parbin:2022iwt}.

\begin{figure}[H]
   \centering

   \begin{subfigure}[b]{0.35\textwidth}
       \centering
       \includegraphics[width=\linewidth]{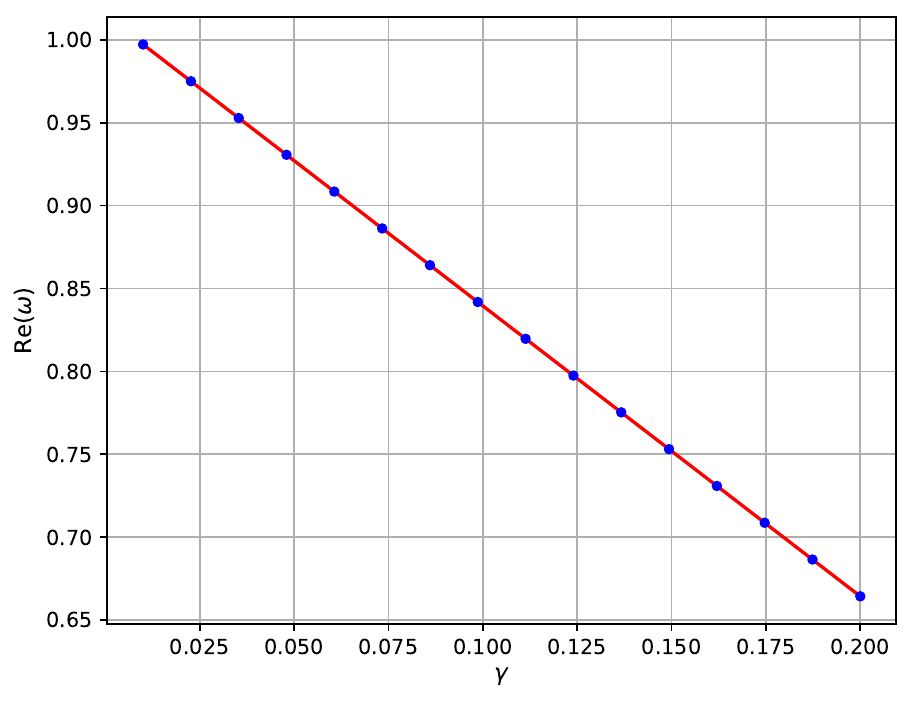}
       \caption{ $\omega_q=-1/3$.}
       \label{fig:ReSigma}
   \end{subfigure}
   \hfill
   \begin{subfigure}[b]{0.35\textwidth}
       \centering
       \includegraphics[width=\linewidth]{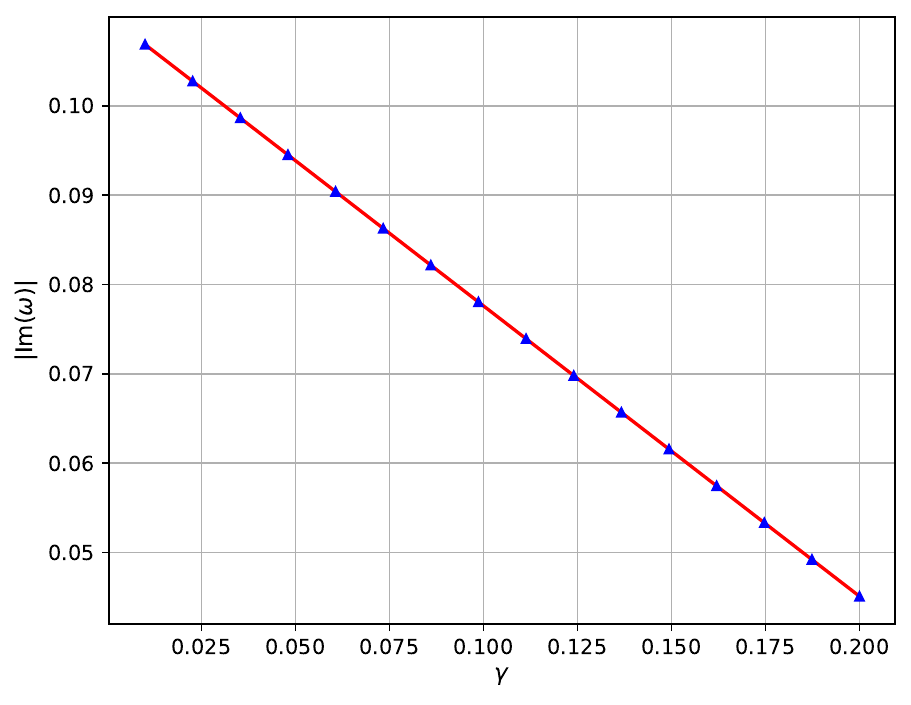}
       \caption{$\omega_q=-1/3$.}
       \label{fig:ImSigma}
   \end{subfigure}

   \vspace{0.35cm}

   \begin{subfigure}[b]{0.35\textwidth}
       \centering
       \includegraphics[width=\linewidth]{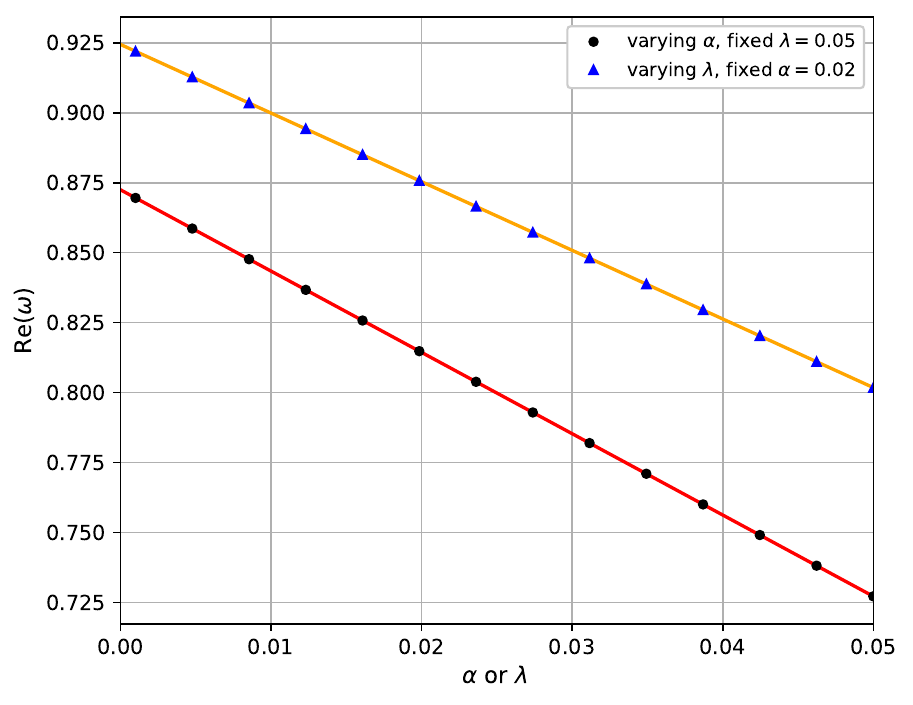}
       \caption{$\omega_q=-2/3$ }
       \label{fig:ReAlphaWq23}
   \end{subfigure}
   \hfill
   \begin{subfigure}[b]{0.35\textwidth}
       \centering
       \includegraphics[width=\linewidth]{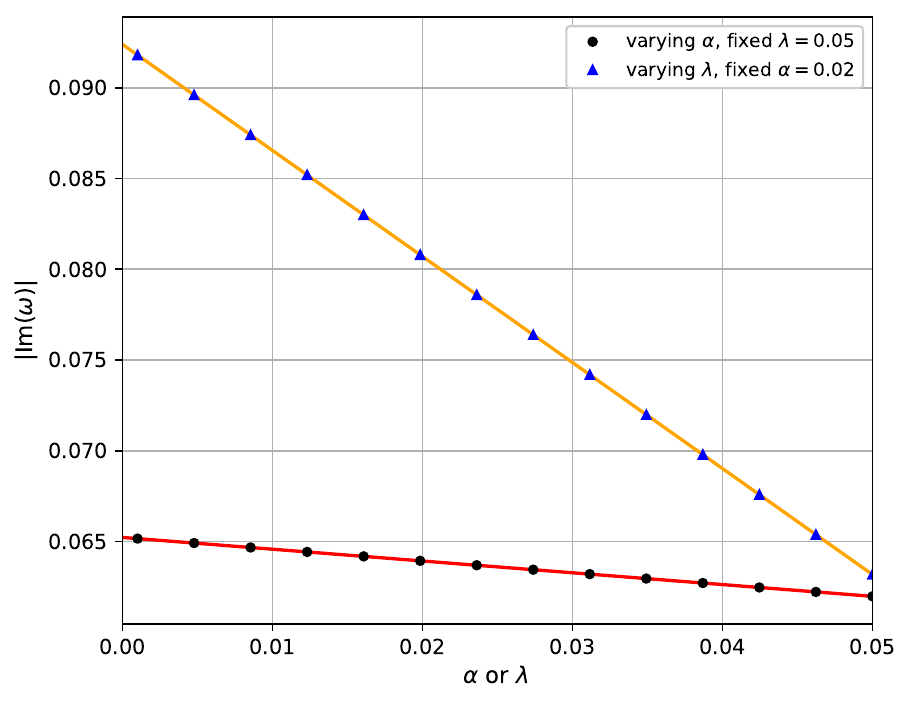}
       \caption{ $\omega_q=-2/3$ }
       \label{fig:ImLambdaWq23}
   \end{subfigure}

   \caption{Variation of the quasinormal-mode frequencies for
   $M=1$, $Q=0.9$, $\ell=4$, and $n=0$.}
   \label{fig:qm}
\end{figure}

The dependence of the QNM frequencies on the exotic-matter parameters
is displayed in Fig.~\ref{fig:qm}. The real part
$\mathrm{Re}(\omega)$ characterizes the oscillation frequency of the
perturbative modes, whereas $|\mathrm{Im}(\omega)|$ measures the
damping rate and therefore controls the relaxation time of the
perturbations.

For the special case $\omega_q=-1/3$, the quintessence contribution
becomes a constant term in the metric function. As a result, the QNM
spectrum does not depend on $\alpha$ and $\lambda$ separately, but only
on the effective combination $\gamma=\lambda+\alpha$. This degeneracy is illustrated in Figs.~\ref{fig:ReSigma} and
\ref{fig:ImSigma}, where both $\mathrm{Re}(\omega)$ and
$|\mathrm{Im}(\omega)|$ are plotted as functions of $\gamma$. Increasing
$\gamma$ modifies the height and width of the effective potential, which
in turn changes both the oscillation frequency and the damping rate of
the scalar perturbations. Therefore, in this limit, the string-cloud and
quintessence contributions cannot be interpreted as two independent
physical effects.

The situation changes for $\omega_q=-2/3$. In this case, one has
$3\omega_q+1=-1$, so that the quintessence contribution behaves as
$-\alpha r$ and becomes radially dependent. Consequently, the
degeneracy with the constant string-cloud term is removed. This is shown
in Figs.~\ref{fig:ReAlphaWq23} and
\ref{fig:ImLambdaWq23}, where the variations of both $\alpha$ and
$\lambda$ lead to distinguishable changes in $\mathrm{Re}(\omega)$ and
$|\mathrm{Im}(\omega)|$. The parameter $\alpha$ mainly reflects the
radial quintessence deformation of the geometry, while $\lambda$
encodes the constant contribution of the string cloud. Their different
radial dependence explains why the two parameters affect the QNM
spectrum differently in the non-degenerate case.

\subsection{Verification of Hod's conjecture}

One of the most important questions in black hole physics is how the fundamental properties, such as mass, charge, and angular momentum of a black hole, are encoded into its quasinormal modes. One can identify a mode whose QNM damping rate is bounded by a universal value proportional to the black hole temperature.
\begin{equation}\label{10}
    |\mathrm{Im}(\omega)| \leq \pi T_H
\end{equation}
where \(T_{H}\) is the Hawking temperature. \\
The Hawking temperature is related to the surface gravity at the event horizon \(r_{+}\), and can be expressed in terms of the metric function \(f(r)\) as \(\kappa=\frac{1}{2}\left.\frac{d f(r)}{dr}\right|_{r=r_{+}}\). Using the relation between surface gravity and temperature, the following equation expresses the Hawking temperature of the black hole in terms of the metric function evaluated at the event horizon. 
\begin{equation}\label{11}
    T_H = \frac{\kappa}{2\pi} = \frac{1}{4\pi r_+} \left[ 1 - \lambda - \frac{Q^2}{r_+^2} + \frac{3\omega_q\alpha}{r_+^{3\omega_q+1}} \right]
\end{equation}

For the particular case \( \omega_q = -\frac{1}{3} \) the expression simplifies to

\begin{equation}
    T_H = \frac{(1 - \lambda - \alpha)^2 \sqrt{M^2 - (1 - \lambda - \alpha) Q^2}}{2\pi \left( M + \sqrt{M^2 - (1 - \lambda - \alpha) Q^2} \right)^2}
\end{equation}

with the event horizon radius given by
\begin{equation}
    f(r_+) = 0 \Rightarrow r_+ = \frac{M + \sqrt{M^2 - (1 - \lambda - \alpha) Q^2}}{1 - \lambda - \alpha}.
\end{equation}

\begin{figure}[H]
    \centering
    \includegraphics[width=1.03\linewidth]{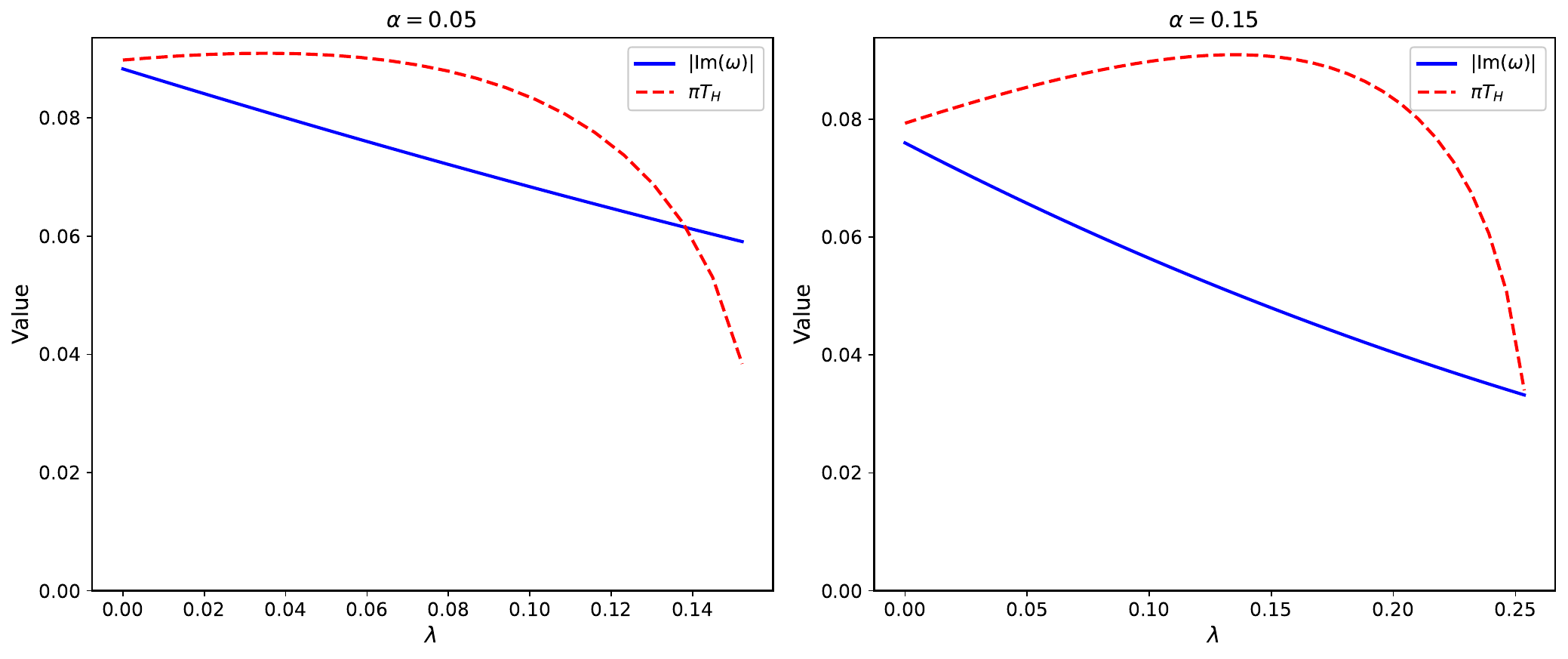}
    \caption{Variation of the magnitude of imaginary quasinormal modes and $\pi T_H$ with respect to model parameter $\lambda$ with $n = 0$ and $\alpha=0.05$, $0.15$}
    \label{fig:placeholder00}
\end{figure}

\begin{figure}[h!]
    \centering
    \includegraphics[width=1.03\linewidth]{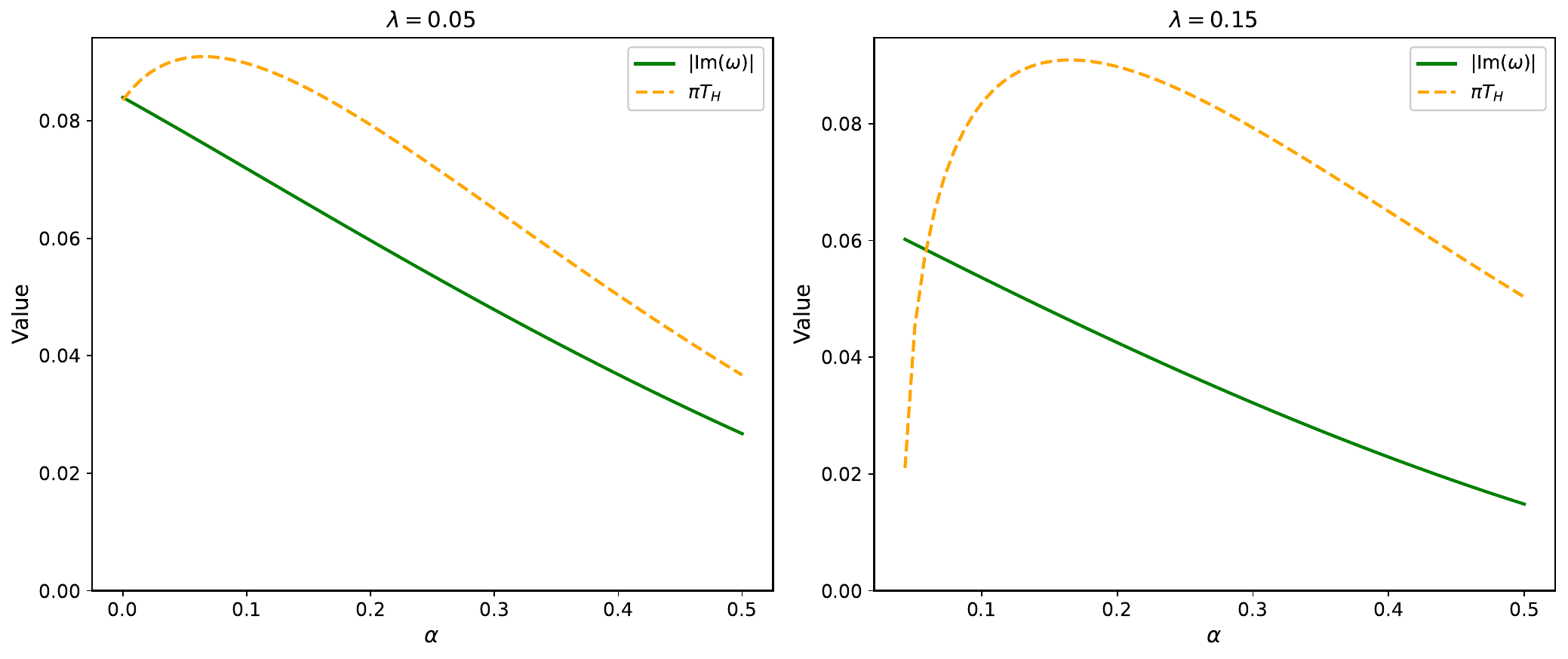}
    \caption{Variation of magnitude of imaginary quasinormal modes and $\pi T_H$ with respect to model parameter $\alpha$ with $n = 0$ and $\lambda=0.05$, $0.15$}
    \label{fig:placeholder11}
\end{figure}

To verify Hod's conjecture, we examine Figs.~\ref{fig:placeholder00}
and~\ref{fig:placeholder11}, where the magnitude of the imaginary part
of the quasinormal frequency, $|\mathrm{Im}(\omega)|$, is compared with
the thermodynamic bound $\pi T_H$. The first figure shows the variation
with respect to the string-cloud parameter $\lambda$ for fixed values of
$\alpha$, while the second shows the variation with respect to the
quintessence parameter $\alpha$ for fixed values of $\lambda$.

For the special value $\omega_q=-1/3$, however, the quintessence
contribution becomes constant and the metric function depends on the
effective combination
$\gamma=\lambda+\alpha$. Consequently, the variations shown in Figs.~\ref{fig:placeholder00}
and~\ref{fig:placeholder11} should be interpreted as different scans of
the same effective parameter $\gamma$, rather than as two independent
physical effects of $\lambda$ and $\alpha$.

The comparison between $|\mathrm{Im}(\omega)|$ and $\pi T_H$ shows that
Hod's bound can be either satisfied or violated depending on the value
of the effective deformation parameter. In the parameter regions where
$|\mathrm{Im}(\omega)|<\pi T_H$, the damping rate remains below the
thermodynamic relaxation bound and the conjecture is respected.
Conversely, when $|\mathrm{Im}(\omega)|$ exceeds $\pi T_H$, the
relaxation rate becomes larger than the allowed thermodynamic limit,
signaling a violation of Hod's bound.

Physically, these results indicate that the exotic matter contribution
modifies both the quasinormal damping rate and the Hawking temperature.
Since, for $\omega_q=-1/3$, the relevant dependence is through
$\gamma=\lambda+\alpha$, the apparent changes induced by varying either
$\lambda$ or $\alpha$ reflect the response of the black hole to the same
effective deformation of the background geometry. Thus, the figures
should not be read as evidence that the string cloud and quintessence
separately stabilize or destabilize the system in this particular
limit.

\section{Optical Properties: Shadow and Emission Rate}
\label{sec:opp}
In this section, we investigate the optical properties of the black hole, focusing on its shadow and the associated emission rate as these observables encode essential information about the spacetime geometry in the strong field regime. Black holes possess extremely strong gravitational fields that profoundly affect the propagation of light. Photons passing sufficiently close to the event horizon may become trapped on unstable circular orbits, leading to the formation of a dark region on the observer’s sky known as the black hole shadow \cite{PhysRevD.110.084016, Borah:2025tvw,Saleh_2011, Zahid:2023csk}. It appears as a silhouette against a bright background of surrounding radiation where light is scattered near the event horizon over a wide range of directions. The shape of this shadow plays a crucial role as an observable target for astrophysical observations and provides a tool to probe strong-field gravity.
 By making use of recent astronomical observations and imaging technologies, we have been able to obtain explicit images of these elusive black hole shadows, thus constituting a large step forward in our understanding of these cosmic phenomena \cite{EventHorizonTelescope:2019dse,EventHorizonTelescope:2022wkp}.

The size and shape of the shadow depend on the underlying spacetime geometry and therefore on the physical parameters of the black hole, such as its mass, spin, electric charge, and possible scalar or exotic matter contributions. Consequently, precise measurements of black hole shadows can be used to constrain deviations from standard Kerr or RN geometries. 
\subsection{Shadow radius}
Here, we explore the correlation between the black hole shadow and the real part of the QNM frequency. To establish this correspondence, we study the shadow of a four-dimensional Reissner-Nordström black hole coupled to a quintessence field and surrounded by a cloud of strings, using the geodesic approach.

The spacetime metric of this black hole solution is spherically symmetric and possesses three Killing vectors, one timelike and two spacelike. These vectors give rise to conserved quantities, namely the energy $E$ and angular momentum $L$. Let $\tau$ denote an affine parameter along the geodesic. The Hamilton-Jacobi equation is

\begin{equation}\label{14}
\frac{\partial S}{\partial\tau} = -\frac{1}{2}g^{\mu\nu}\frac{\partial S}{\partial x^{\mu}}\frac{\partial S}{\partial x^{\nu}},
\end{equation}
where $S$ is the Jacobi action. For the RN BH with quintessence and a cloud of strings, the Jacobi action can be decomposed as
\begin{equation}\label{15}
S = \frac{1}{2}m_{0}^{2}\tau - Et + L\phi + S_{r}(r) + S_{\theta}(\theta),
\end{equation}
where $m_0$ is the mass of the test particle (zero for photons), $E$ is the conserved energy, and $L_\phi$ is the conserved angular momentum in the $\phi$ direction. The functions $S_{r}(r)$ and $S_{\theta}(\theta)$ only depend on the radial and polar coordinates, respectively.

From these conserved quantities and their conjugate relationships, we obtain the geodesic equations
\begin{equation}
       \begin{cases} \dfrac{dt}{d\tau} = \dfrac{E}{f(r)} ,\\
       \dfrac{d\varphi}{d\tau} = \dfrac{L}{r^2 \sin^2{\theta}}. 
    \end{cases}
     \end{equation}

Inserting Eq.~\eqref{14} into Eq.~\eqref{15} and performing various calculations, we can write out the separated equations of radial and angular as follows
\begin{equation}
       \begin{cases} 
        \dfrac{dr}{d\tau} = \pm \sqrt{\mathcal{R}(r)} \\
       r^2 \dfrac{d\theta}{d\tau} = \pm \sqrt{\Theta(\theta)},
    \end{cases}
     \end{equation}
the terms $\mathcal{R}$ and $\Theta$ are expressed as
\begin{equation}\label{3.13}
    \mathcal{R}(r) = E^2 \left(1 - \frac{f(r)}{r^2} (\eta + \zeta^2) \right),
\end{equation}
\begin{equation}
\Theta(\theta) = E^{2}\left( \eta - \zeta^{2}\cot^{2}\theta \right)
\end{equation}
with 
\begin{equation}\label{3.12}
    \zeta = L/E, \quad \eta = \mathcal{K}/E^2,
\end{equation}
where $\mathcal{K}$ represents the Carter constant (or separated angular constant).

In the observer's image plane, we define celestial coordinates in terms of ratios (the factor r guarantees a finite scale when \( r_0 \to \infty \)) 

\begin{equation}
x = -\lim_{r_0 \to \infty} \left( r_0\frac{p_{\phi}}{p_{t}} \right), \quad y = \lim_{r_0 \to \infty} \left( r_0 \frac{p_{\theta}}{p_{t}} \right).
\end{equation}

Then, by substituting the momentum components, we immediately obtain 

\begin{equation}
x = -\frac{L}{E \sin \theta_0}, \quad y = \frac{p_\theta}{E},
\end{equation}
where $(\theta_0)$ is the observer inclination angle, and $p_\theta$ is evaluated along the geodesic at the observer’s position.

 $p_i$ $p_\mu$ are the covariant components of the four-momentum. These contravariant components can be calculated as \cite{Hioki_2009,Raza:2025ohk}
\begin{equation}
     p_\mu=\frac{\partial \mathcal{L}}{\partial\dot{x^\mu}}=g_{\mu\nu}\dot{x^\nu} \quad \text{and}\quad \mathcal{L}=\frac{1}{2}g_{\mu\nu}\dot{x}^\mu\dot{x}^\nu ,
\end{equation}
Here $\zeta$ and $\eta$ are the impact parameters defined as equation~\eqref{3.12}, where for null geodesics, one obtains the following function to separate variables
\begin{equation}
\Theta(\theta) = p_\theta^2 = E^2 \left( \eta - \zeta^2 \cot^2 \theta \right).
\end{equation}
Thus, for an observer at angle $\theta_0$, we find
\begin{equation}
p_\theta = \pm E \sqrt{\eta - \zeta^2 \cot^2 \theta_0}.
\end{equation}
Substituting into the definitions of $x$ and $y$ yields the general expressions

\begin{equation}\label{27}
x = -\frac{\zeta}{\sin\theta_0}, \quad y = \pm\sqrt{\eta - \zeta^2 \cot^2\theta_0}.
\end{equation}

The $( \pm )$ in $y$ match the two symmetric directions with respect to the image axis.

In the equatorial plane $( \theta_0 = \frac{\pi}{2} )$ for an observer we have
 \begin{equation}
     x = -\zeta, \quad y = \pm\sqrt{\eta}.
 \end{equation}
The fundamental connection between celestial coordinates and impact parameters, particularly relevant for equatorial observers ($\theta_0 = \pi/2$), is mathematically expressed in Eq.~\eqref{27}.

We further establish a connection between the impact parameters of the metric~\eqref{eq:f(r)} by utilizing the conditions for unstable spherical photon orbits $(\mathcal{R} = 0)$ and $(d\mathcal{R}/dr = 0)$, resulting in

\begin{equation}
r = \left. \frac{2f(r)}{f'(r)} \right|_{r = r_{ph}},
\end{equation}

and

\begin{equation}\label{29}
\eta + \zeta^2 = \left. \frac{r^2}{f(r)} \right|_{r = r_{ph}},
\end{equation}

\begin{equation}\label{30}
 \frac{r_{ph}^2}{f(r_{ph})} = R_s^2.
\end{equation}
Where \(R_s\) represents the shadow radius of the black hole. The photon sphere \(r_{ph}\) characterizes the critical orbits that define the boundary of the black hole shadow.

\begin{figure}[H]
    \centering
    \includegraphics[width=1.09\linewidth]{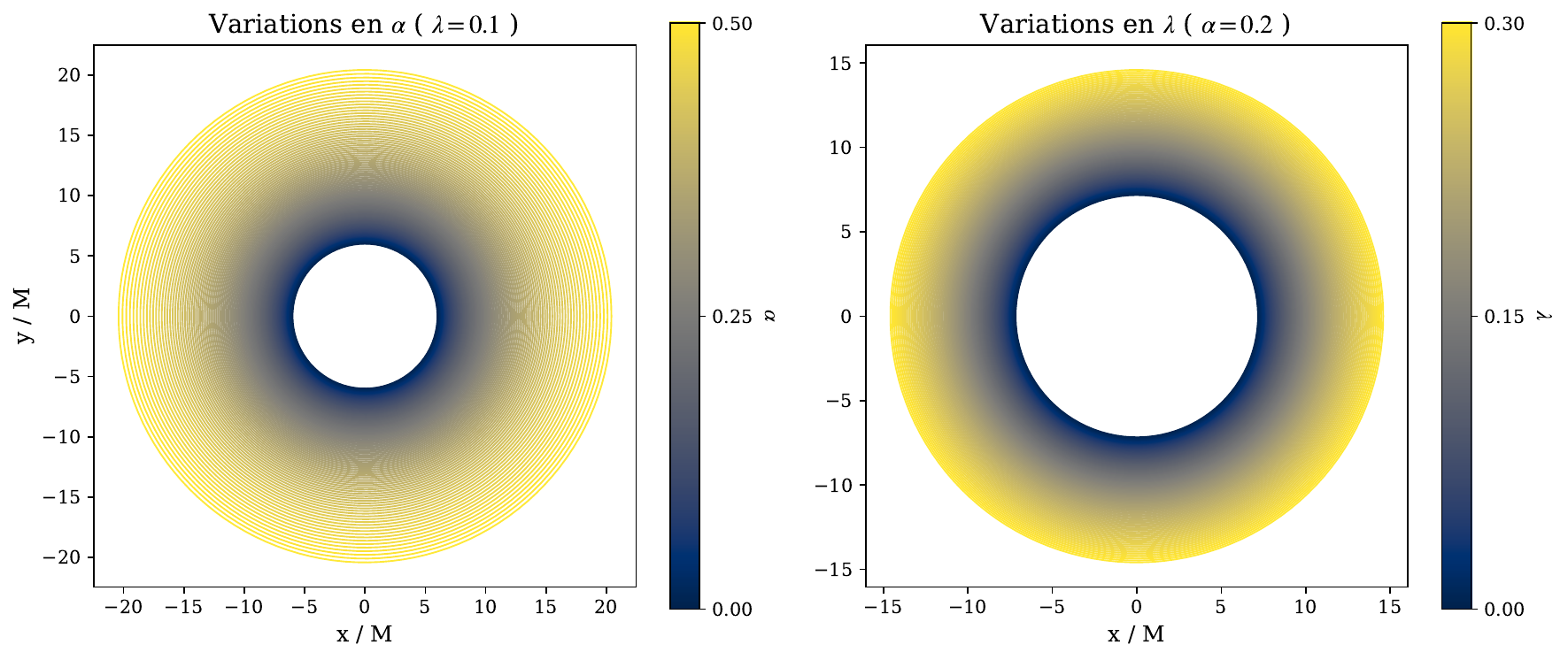}
    \caption{  Black hole shadow dependence on quintessence ($\alpha$) and the cloud of string ($\lambda$). 
            Left: Variation with $\alpha$ ($\lambda = 0.1$ fixed). 
            Right: Variation with $\lambda$ ($\alpha = 0.2$ fixed). 
            The color gradient shows parameter progression from zero (dark) to maximum values (yellow).}
    \label{shad}
\end{figure}

In figure \ref{shad} we have set up a detailed analysis of the size of the shadow in terms of $\alpha$ and $\lambda$. The shadow appears as a dark region on the observer’s sky, surrounded by a bright photon ring formed by light rays that asymptotically approach the unstable photon sphere.

The left panel shows the variation of the shadow with increasing quintessence density $\alpha$ for fixed string cloud density $\lambda = 0.1$. The gradual expansion of the shadow radius with $\alpha$ demonstrates that quintessence acts as a cosmological field that alters the asymptotic spacetime geometry. This behavior originates from the added $-\alpha/r^{3\omega_q+1}$ term in the metric function $f(r)$, leading to a deeper gravitational potential well. In the physical phase of quintessence, it makes gravitational focusing stronger and raises the critical impact parameter for photon capture. Even moderate values $\alpha\sim 0.1-0.2$ lead to a noticeable enlargement of the shadow compared to the vacuum RN case.

The right panel displays the dependence on the string cloud parameter for fixed $\alpha=0.2$. The linear $\lambda$ term in the metric function corresponds to a homogeneous energy density permeating spacetime, which increases the gravitational mass measured at infinity, thus enlarging the photon sphere radius and the shadow size. Consequently, the presence of a string cloud increases the black hole’s gravitational cross section, acting analogously to an effective mass enhancement while arising from a distinct stress–energy tensor structure.
\subsection{Study of the violation of Hod's conjecture through the optical properties of black holes}
Motivated by these optical properties discussed in the previous subsection, we now establish the connection between the BH shadow and the imaginary quasi-normal modes. As already observed in the eikonal limit, the relationship between black hole shadow and real quasi-normal modes of a black hole has been studied  \cite{Cuadros-Melgar:2020kqn}. In the present analysis, however, we focus on the damping rate of the quasi-normal modes and investigate how optical properties influence the validity of Hod’s conjecture. For mathematical convenience, we consider the expansion of the quasinormal mode frequency truncated at third order
\begin{equation}\label{35}
\begin{aligned}
\omega &= \left\{V_s + \frac{V_4}{8V_2}\left(\nu^2 + \frac{1}{4}\right) - \left(\frac{7 + 60\nu^2}{288}\right)\frac{V_3^2}{V_2^2}\right. \\
&\quad + i\nu\sqrt{-2V_2}\left[\frac{1}{2V_2}\left[\frac{5V_3^4(77 + 188\nu^2)}{6912V_2^4} \right.\right. \\
&\quad \left.\left.\left.- \frac{V_3^2V_4(51 + 100\nu^2)}{384V_2^3}
 + \frac{V_4^2(67 + 68\nu^2)}{2304V_2^2}\right.\right. \right. \\  
 &\quad \left.\left.\left.+ \frac{V_5V_3(19 + 28\nu^2)}{288V_2^2}
+ \frac{V_6(5 + 4\nu^2)}{288V_2}\right] - 1\right]\right\}_{r=r_{\text{ph}}}^{1/2}.
\end{aligned}
\end{equation}
In this expression, \( V_j \) denotes the $j^{th}$-derivative of the potential \( V_s \) with \( \nu = n + \frac{1}{2} \). The third order formula does not affect any results at the eikonal limit, since it yields the identical expansion as for the first terms of this expansion, as given by representations of orders ( \( 4^{\text{th}} \) to \( 6^{\text{th}} \) ). The quasi-normal frequency \( \omega \)  has been computed at the radius \( r_0 \) which corresponds to the maximum value of the potential \( V \), determined in \cite{Gogoi:2024vcx} via the following condition

\begin{equation}
    \frac{dV}{dr}\Big|_{r=r_0}=f(r)\left( \frac{\ell(\ell+1)}{r^2} + \frac{f'(r)}{r} \right)\Big|_{r=r_0}=0
\end{equation}
This condition determines the location \( r_0 \) of the maximum of the effective potential, which in general depends on the black hole parameters, the matter content, and the underlying gravitational theory. However, in the eikonal limit (\(\ell \gg 1\)), where the angular momentum term dominates, the leading and first subleading contributions simplify this condition, yielding the universal relation

\begin{equation}
\left.\frac{d}{dr}\!\left(\frac{f(r)}{r^2}\right)\right|_{r=r_0}=0.
\end{equation}
Defining \(G=\dfrac{f(r)}{r^2}\), this equation can be written in the form
\begin{equation}
\left.\frac{d}{dr} \! \bigl(G\bigr)\right|_{r=r_0}=0,
\qquad
\left.\frac{d}{dr}\!\bigl(G^{-1}\bigr)\right|_{r=r_{ps}}=0.
\end{equation}

As a consequence, one finds that \(r_{ps}=r_0\), provided that \(G^{-2}|_{r=r_{ps}}\neq0\), which is the case in general.  We obtain at the eikonal regime, as a result \(r_{ps}=r_0\) provided \(G^{-2}|_{r=r_{ps}}\neq0\), which is generally the case.

This result demonstrates a universal property of spherically symmetric black holes: the position of the maximum
of the potential of motion equations of fields corresponds to the stability threshold for the circular null geodesic around the
structure.

Finally, by expanding the relation \ref{35} and considering the real part only,
Expanding relation \ref{35} and keeping the real part only, we find at eikonal order

\begin{equation}
\omega_R = \frac{1}{2} R_s^{-1}\bigl(\ell + 1 + \mathcal{O}(\ell^{-1})\bigr).
\end{equation}
This relation connects the shadow and the quasinormal frequency of the black hole. Moreover, the imaginary terms in the
eikonal regime,
\begin{equation}
\omega_I = \frac{n+1}{2\sqrt{2}}\,R_s^{-1}\,\sqrt{\,2f - r^2 f''\,}\Bigg|_{r=r_{ps}} \;+\; \mathcal{O}(\ell^{-1}),
\end{equation}
indicate that the factor 
\(\sqrt{\,2f - r^2 f''\,}\big|_{r=r_{ps}}\) directly controls the damping rate of the quasinormal modes and thus governs the degree to which the black hole's optical properties can lead to deviations from Hod's conjecture.

\subsection{Emission rate}
The particle emission rate of a black hole is another important aspect of its optical properties. And can provide a useful probe of its stability and lifetime. It is primarily associated with two important properties of the black hole: its Hawking temperature and its shadow. The energy emission rate is as follows \cite{Wei_2013,Page:1976df,Al-Badawi:2024mco,Decanini:2011xi,Cuadros-Melgar:2020kqn}
\begin{equation}\label{31}
     \frac{d^2 E(\bar{\omega})}{d\bar{\omega} dt} = \frac{2\pi^2 \sigma_{\text{lim}}}{\exp\left( \frac{\bar{\omega}}{T_H} \right) - 1} \bar{\omega}^3 ,
\end{equation}
where $E(\bar{\omega})$ and $\bar{\omega}$ stand for the emitted energy and the radiation frequency.

The expression for $\sigma_{\text{lim}}$, which is the limiting constant value, is expressed in $D$ spacetime dimensions as 
\begin{equation}
    \sigma_{\text{lim}} = \frac{\pi^{\frac{D-2}{2}} R_s^{D-2}}{\Gamma \left( \frac{D}{2} \right)},
\end{equation}
For \(D = 4\); it reduces to:
    \begin{equation}
        \sigma_{\text{lim}}  = \pi R_s^2,
    \end{equation}
    
Substituting this in \ref{31}, the energy emission rate become
  \begin{equation}\label{34}
      \frac{d^2 E(\bar{\omega})}{d t d \bar{\omega}}  = \frac{2\pi^3 R_s^2 \bar{\omega}^3}{\exp\left( \frac{\bar{\omega}}{T_H} \right) - 1}.
  \end{equation}

\begin{figure}[H]
   \centering
   \begin{subfigure}{0.4\textwidth}
       \includegraphics[width=\linewidth]{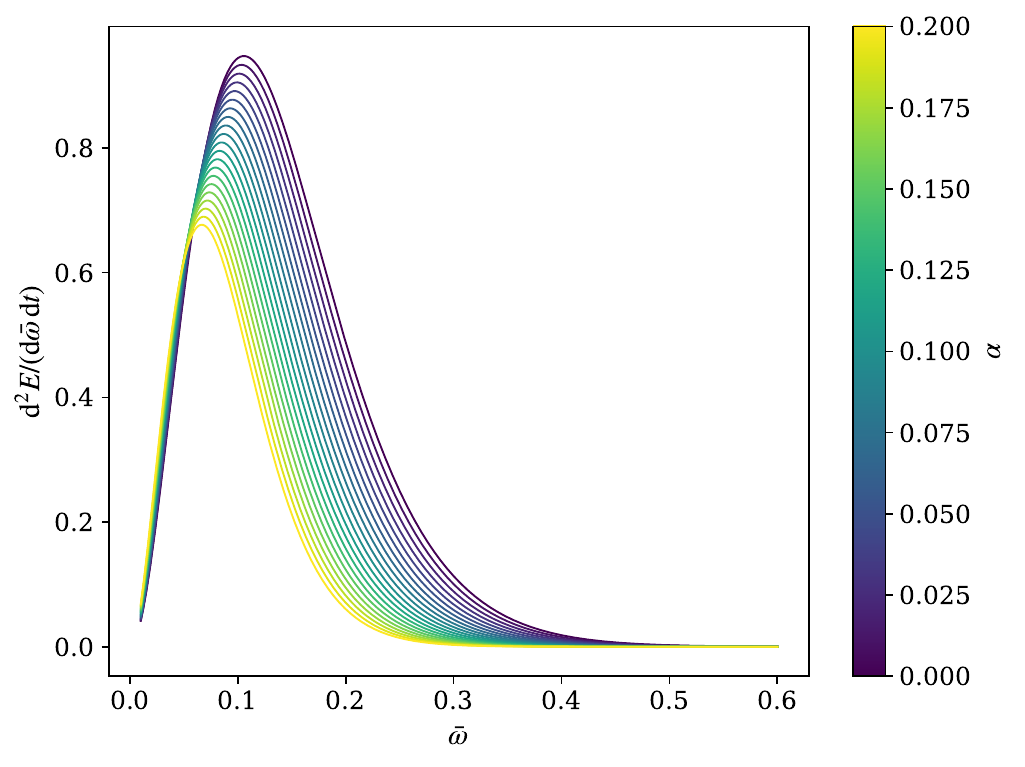}
       \caption{ $\lambda=0.03$}
       \label{rea}
   \end{subfigure}
   \hfill
   \begin{subfigure}{0.4\textwidth}
       \includegraphics[width=\linewidth]{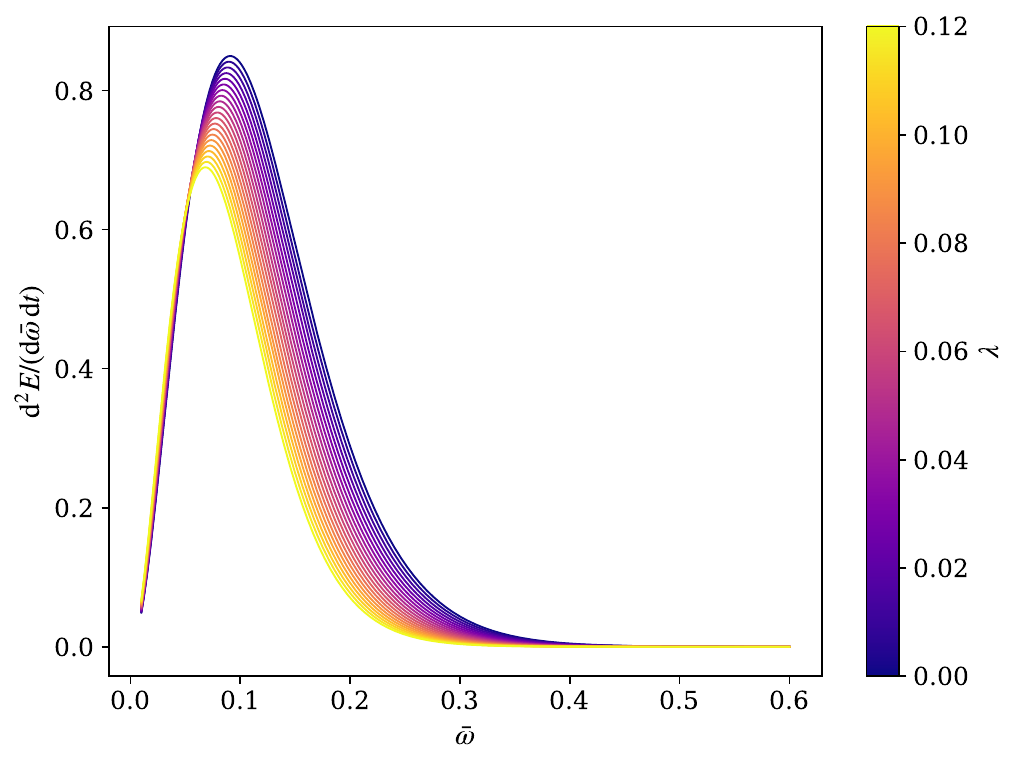}
       \caption{ $\alpha=0.1$ }
       \label{ima}
   \end{subfigure}
\caption{Energy emission rate for fixed  $M=1$, $Q=0.3$ and $\omega_q=-1/3$}
   \label{fig:ems}
\end{figure}

In Fig. \ref{fig:ems} we show the energy emission rate of an RN BH coupled to a quintessence field and surrounded by a string cloud, for $M=1$ and  $Q = 0.3$, $\omega_q=-1/3$. It exhibits the well-known bell-shaped form that is commonly associated with black hole radiation
emission is suppressed at low frequencies, rises to a peak, and is exponentially damped at high frequencies.

This expression \ref{34} shows that the emission spectrum is controlled by two competing factors, namely the cubic dependence of frequency $\bar\omega^3$, which is the dominant term at lower frequencies, and an exponential suppression factor $\exp(\bar{\omega}/T_H)-1$, which comes into play at higher frequencies. 

Such a spectrum of emitted radiation is altered due to the existence of exotic matter fields that affect both the shadow radius $R_s$ and the temperature $T_H$ of the emitted Hawking radiation. With parameter $\alpha$, a gravitational screening appears for quintessence, which lowers the Hawking temperature. This thermal quenching leads to a blue shift of the emission peak, accompanied by an overall reduction in intensity. Meanwhile, the string cloud parameter $\lambda$ provides a constant energy density term which leads to a decreased value for the shadow radius and effectively enlarges the gravitational capture cross-section as well as boosts the low-frequency emission.

The cumulative effect of all these exotic matter fields is a wider and less intense emission spectrum than in the standard Reissner-Nordström case. This altered radiative character has important implications for observational features of black holes surrounded by exotic matter fields. The suppressed emission rates indicate such black holes would have a longer evaporation timescale and potentially different electromagnetic observables in astrophysical settings.

\section{Hod–Swampland Distance Conjecture Bound} 
\label{sec:hsdb}
The Swampland Distance Conjecture (SDC) is a key ingredient of the recent research program to carve out a swampland from the landscape of consistent quantum gravity theories. It conjectures that any finite displacement in the moduli space of a scalar field at a scale greater than the Planck mass induces an infinite tower of exponentially light states signaling the breakdown of effective field theory. In cosmology, this conjecture has important implications for quintessence models, where a scalar field drives the dark energy through its low-energy potential and varies on cosmological timescales and spatial scales \cite{Storm:2020gtv,Brahma:2019kch,Cicoli:2021skd,Gogoi:2024scc}.

In what follows, we examine the interconnection between the Swampland Distance Conjecture and the Hod bound, a template that restricts the damping rate of quasinormal modes and relates the relaxation behavior of black holes to elementary constraints in quantum gravity.
From a black hole to its Hawking temperature.
\subsection{Modeling the quintessential field}

The quintessence can alternatively be defined as a scalar field associated with an exponential potential of the form \cite{Andriot:2025los,Tsujikawa:2013fta}
\begin{equation}
    V_Q(\phi) = V_0 e^{-\beta \phi / M_{\rm Pl}},
\end{equation}
where $\beta$ is a dimensionless parameter and $M_{\rm Pl}$ represents Planck's mass.

The fundamental relationship describing the evolution of the energy density of a
scalar field in an expanding Friedmann-Lema\^itre-Robertson-Walker (FLRW) universe
is given by
\begin{equation}\label{42}
    \rho_{\phi} \propto a^{-3(1+\omega_q)},
\end{equation}
where $a(t)$ denotes the cosmological scale factor. To transpose this behavior to the static, spherically symmetric environment of a black hole \cite{Liu:2025dhl,Kiselev:2002dx,Copeland:1997et}, a heuristic analogy is established between the dynamic role of the scale factor \( a(t) \) in cosmology and that of the radial coordinate \( r \) in static spacetime \cite{Caldwell:1997ii}. This approach is motivated by the fact that in both cases, these parameters govern the dilution or variation in intensity of fields and energy densities with the scale of space-time. Accordingly, we implement the substitution $a(t) \longrightarrow r$ \cite{Nekouee:2023tew}, which leads us to
\begin{equation}\label{43}
    \rho_{\phi}(r) \propto r^{-3(1+\omega_q)}.
\end{equation}
This formulation is not arbitrary. It relies on a well-established correspondence between the physics of scalar fields in expanding cosmological backgrounds and in static gravitational configurations \cite{NooriGashti:2024bbc,Sun:2022wya}. In cosmology,  \( a(t) \) controls the dilution of energy density due to expansion, while in black hole spacetimes, the radial coordinate r controls the spatial decay of energy distributions away from the central singularity. This approach has been successfully employed in a number of works to construct black hole solutions surrounded by quintessence or generic scalar fields, where Eqs.~\ref{42} and \ref{43} are imposed as ansatz for solving the Einstein equations.

In the context of exponential potentials characteristic of quintessence models, scaling solutions emerge where the scalar field exhibits a logarithmic radial dependence,
\begin{equation}
    \phi(r) \propto \ln\!\left(\frac{r}{r_{\rm ref}}\right),
\end{equation}
\cite{Copeland:1997et,Ferreira:1997hj,Garfinkle:1990qj}, reflecting the self-similar behavior of the field configuration under radial transformations and maintaining a consistent energy density profile across different spacetime scales.
This functional form of the potential
\begin{equation}
    V_Q(\phi(r)) \propto r^{-3(1+\omega_q)}
\end{equation}
is exactly what follows from inserting the logarithmic profile for a field into an exponential potential, which connects the scalar dynamics and the geometry that results \cite{Ferreira:1997hj,Gibbons:1987ps,Barreiro:1999zs}. That the radial dependence of these power laws is just the form which would have made the quintessence energy density to decrease as \ref{43} (that is, appropriately for keeping this away from 1 as we go around in empty space), and ensures remains constant that the value of \( \omega_q\) will continue unchanged throughout spacetime.

More concretely, we have the radial profile of the scalar field given by
\begin{equation}
    \phi(r)
    =
    \frac{3(1+\omega_q)M_{\rm Pl}}{\beta}
    \ln\!\left(\frac{r}{r_{\rm ref}}\right).
\end{equation}

\subsection{Field excursion and Hawking temperature}

The connection between scalar field dynamics and black hole thermodynamics becomes particularly meaningful when embedded in a cosmological setting. In a FLRW universe dominated by quintessence, the Hubble horizon \(r_{\text{Hubble}} = H^{-1}\) defines the fundamental scale beyond which sources are no longer causally connected to the black hole  \cite{PhysRevD.15.2738,Cadoni:2024rri}. This, therefore, provides a natural reference scale for defining the scalar field excursion.

We model the black hole as embedded within a spherical region of radius \(r_{\rm ref} = r_{\text{Hubble}}\), and normalize the scalar field such that
\begin{equation}
    \phi(r_{\text{Hubble}}) = 0.
\end{equation}

The field excursion between the Hubble horizon and the black hole event horizon \(r_+\) is defined as:
\begin{equation}
\begin{aligned}
    \Delta\phi
    &= \phi(r_{\text{Hubble}}) - \phi(r_+)  \\
    &= -A \ln\!\left(\frac{r_+}{r_{\rm ref}}\right),
    \qquad
    A = \frac{3(1+\omega_q)M_{\rm Pl}}{\beta}.
\end{aligned}
\end{equation}

This relation can be inverted to express the horizon radius in terms of the field excursion:
\begin{equation}\label{47}
    r_+ = r_{\rm ref} e^{-\Delta\phi/A}.
\end{equation}

The Hawking temperature is given by \ref{11}, substituting the horizon radius Eq.~\ref{47} \cite{Ooguri:2006in,Kehagias2019ANO}, yields the Hawking temperature as an explicit function of the field excursion:
\begin{equation}
\begin{aligned}
 T_H(\Delta\phi)
 &=
 \frac{e^{\Delta\phi/A}}{4\pi r_{\rm ref}}
 \biggl[
 1-\lambda
 -\frac{Q^2}{r_{\rm ref}^2}e^{2\Delta\phi/A}\\
 &+\frac{3\omega_q\alpha}{r_{\rm ref}^{3\omega_q+1}}
 e^{(3\omega_q+1)\Delta\phi/A}
 \biggr].
\end{aligned}
\end{equation}
where we have absorbed the factor \(r_{\text{Hubble}}\) into the coefficients by appropriate scaling of parameters.

\subsection{Modified Hod bound and consistency constraints}

In our work, we adopt the Swampland Distance Conjecture (SDC), which can be stated as follows
\begin{equation}
    |\Delta\phi| \lesssim \mathcal{O}(1) M_{\rm Pl},
\end{equation}
which ensures the validity of our effective description for the quintessence field surrounding the black hole.

When combined with the original Hod conjecture Eq.~\ref{10}, which establishes a universal bound on quasinormal mode damping rates, we derive a modified Hod bound that incorporates both the exotic matter content and quantum gravity constraints \cite{Hod:2006jw,Kiselev:2002dx,Alipour:2023css}. The generalized bound takes the form:
\begin{equation}
\begin{aligned}
 |\mathrm{Im}(\omega)|
 &\leq
 \frac{e^{\Delta\phi/A}}{4 r_{\rm ref}}
 \biggl[
 1-\lambda
 -\frac{Q^2}{r_{\rm ref}^2}e^{2\Delta\phi/A}\\
 &+\frac{3\omega_q\alpha}{r_{\rm ref}^{3\omega_q+1}}
 e^{(3\omega_q+1)\Delta\phi/A}
 \biggr].
\end{aligned}
\end{equation}

This modified upper bound reflects some important physical consequences. First, the bound has an implicit dependence on the string cloud parameter $\lambda$ through the event horizon radius $r_+$ and hence on the field excursion $\Delta\phi$. Second, the quintessence equation of state parameter $\omega_q$ directly scales the exponential sensitivity to field distance. As $\Delta\phi$ approaches the Planck scale, the bound becomes increasingly restrictive, reflecting the effects anticipated by the SDC.

The consistency between these two conjectures is not trivial and puts constraints on the parameter space of our black hole model. In particular, for the resulting combinations of $\alpha$, $\lambda$, and $\omega_q$ that lead either to a violation of the SDC or to an excess of the quasinormal mode damping rate beyond the modified Hod bound, must be excluded from the physically viable regime.

\begin{figure}[ht]
  \centering
  \begin{subfigure}[b]{0.32\textwidth}
    \includegraphics[width=\linewidth]{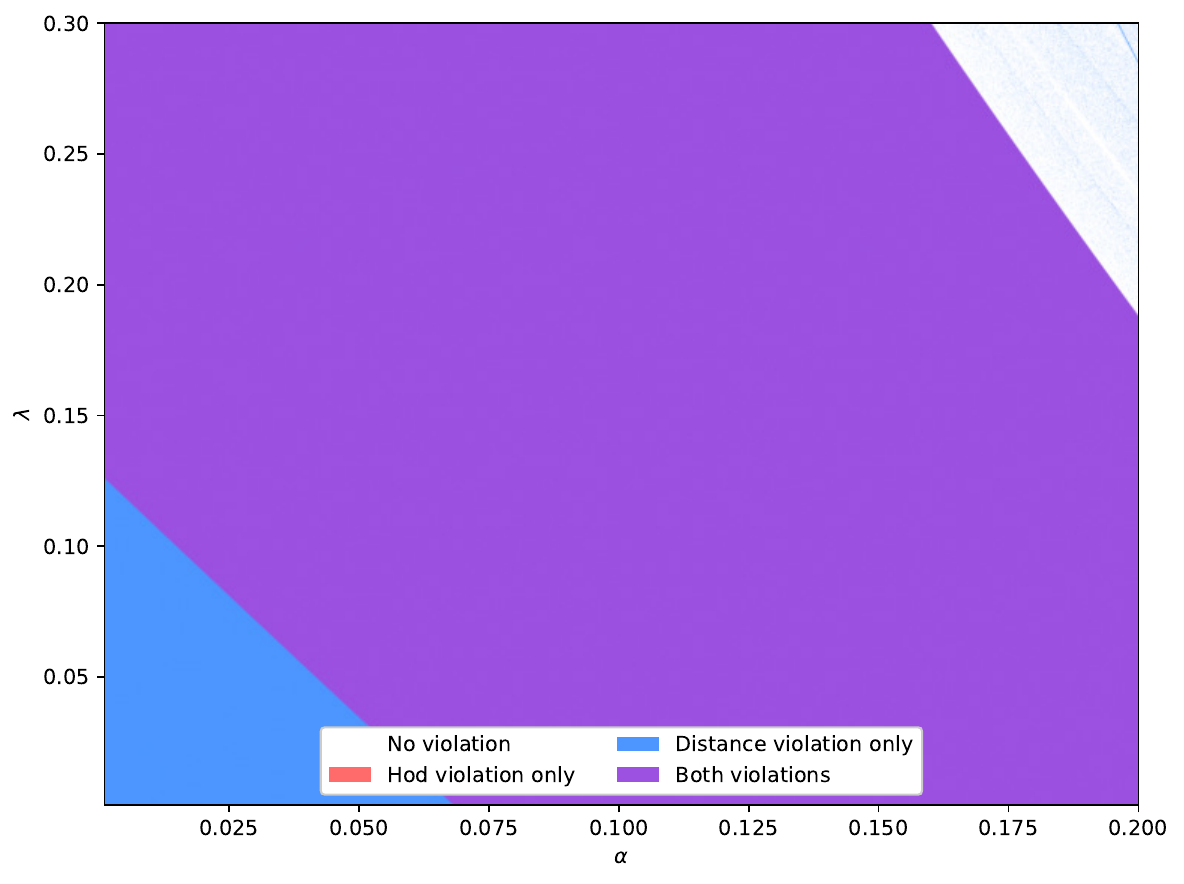}
    \caption{$\omega_q = -1/2$}
    \label{fig:omega_1_2}
  \end{subfigure}
  \hfill
  \begin{subfigure}[b]{0.32\textwidth}
    \centering
    \includegraphics[width=\linewidth]{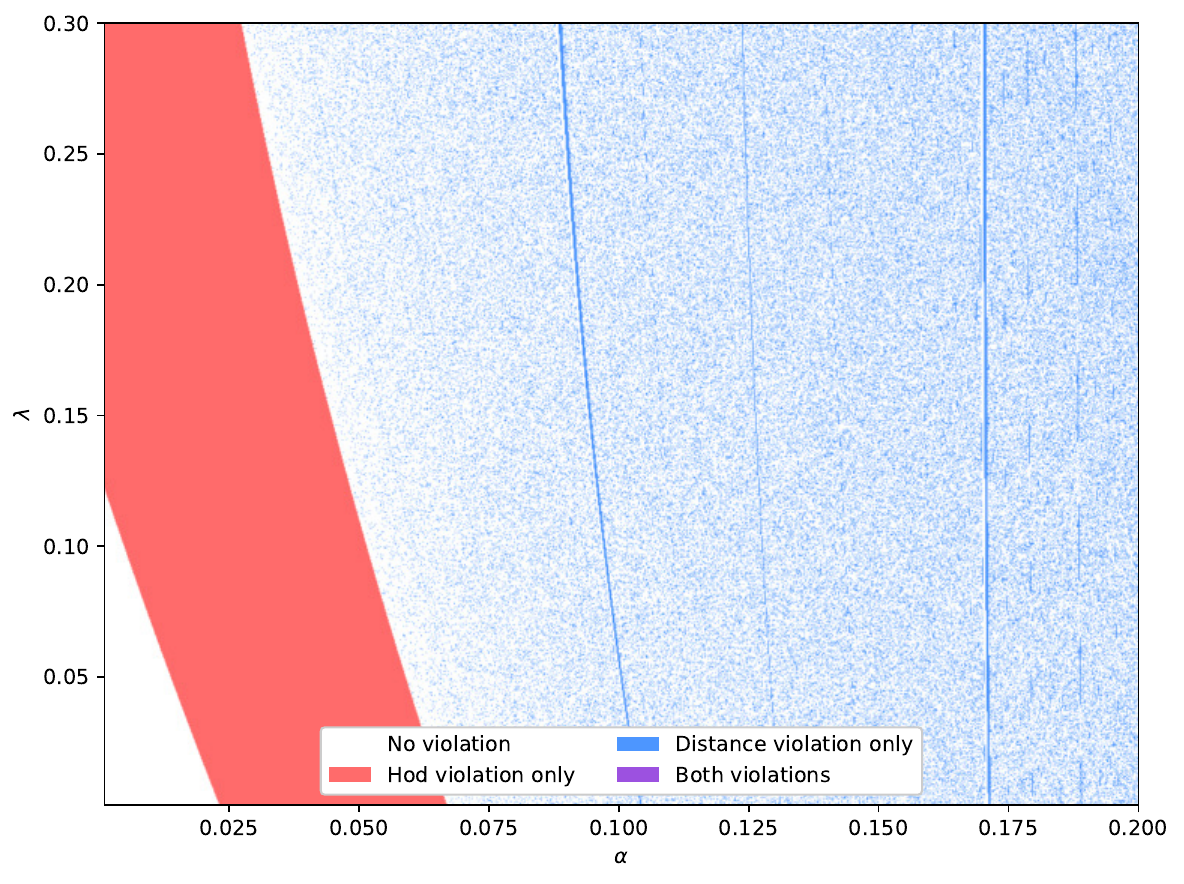}
    \caption{$\omega_q = -5/6$}
    \label{fig:omega_5_6}
  \end{subfigure}
  \hfill
  \begin{subfigure}[b]{0.32\textwidth}
    \centering
    \includegraphics[width=\linewidth]{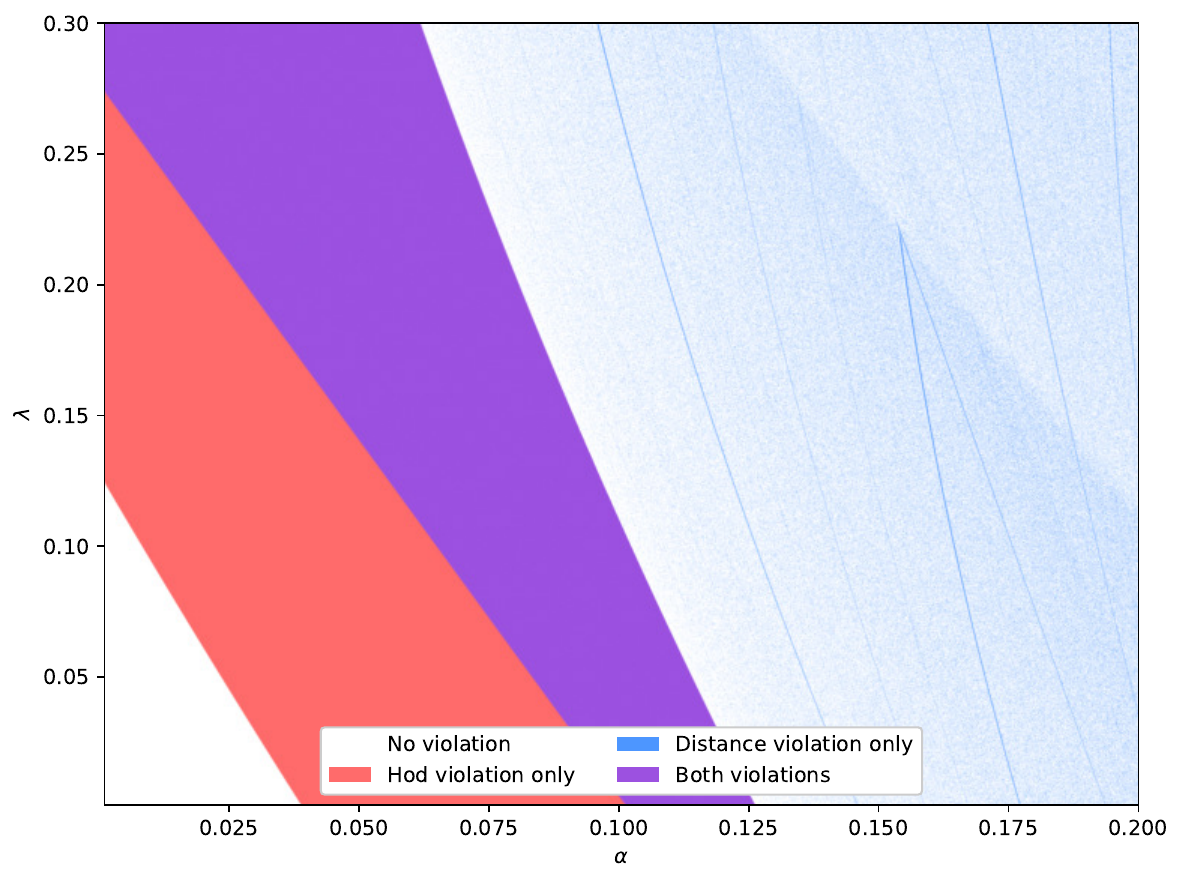}
    \caption{$\omega_q = -2/3$}
    \label{fig:omega_2_3}
  \end{subfigure}
 
  \caption{Allowed regions for Hod and SDC with $M=1$, $Q=0.3$ and $\beta= 1$.}
  \label{fig:vio1}
\end{figure}

The plots in Fig. \ref{fig:vio1} show the interplay between the Hod bound and different limits of the Swampland Distance Conjecture (SDC) for a black hole with quintessence and seated inside a string cloud, parameterized by $\alpha$ and $\lambda$, respectively. Each colored region is associated with a different physical regime: in the `no violation' area, there is overlap between configurations where both conjectures hold simultaneously, and thus thermodynamic consistency coexists with quantum-gravity validity. The Hod-violation-only region corresponds to configurations in which the quasinormal mode decay rate exceeds the modified Hod bound, signaling a breakdown of the universal relaxation constraint despite the validity of the scalar effective theory. Conversely, the SDC-violation-only region indicates field excursions beyond the Planck scale, where quantum gravity effects invalidate the effective scalar description even though the black hole relaxation dynamics remain semiclassical. Finally, the both-violations region corresponds to a complete breakdown of the semiclassical framework.
Physically, these diagrams show how the interplay between scalar-field dynamics and black-hole thermodynamics constrains the viable space of parameters in theories consistent with quantum gravity. Larger values of $\alpha$ strengthen the quintessence field, deepening the gravitational potential well and increasing the scalar field excursion, while larger values of 
$\lambda$ enhances the gravitational contribution of the string cloud, modifying the horizon structure and relaxation dynamics. Excessively strong couplings inevitably drive the system beyond the regime of semiclassical validity.

Remaining within the allowed region ensures that both the relaxation of black hole perturbations and the evolution of the scalar field can be consistently described within an effective field theory framework. Outside this region, either thermodynamic stability or quantum gravity consistency is lost. These bounds therefore provide a sharp diagnostic for identifying regimes where classical black hole physics remains trustworthy and those where quantum gravity effects necessarily dominate. This relationship between black hole phenomenology and quantum gravity constraints furnishes us with a heavy-handed probe of the credibility of exotic matter scenarios in astrophysical settings and an exploration of the limits of applicability of effective field theory when taken into strong gravity regimes.

\section{Conclusions} \label{sec:conclusions}
In this work, we have presented a comprehensive investigation of the dynamical and optical properties of an RN black hole surrounded by both a quintessence field and a cloud of strings. The methodological approach was based on the computation of quasi-normal modes using the higher-order WKB method with Padé averaging, providing a precise characterization of the black hole's response to scalar perturbations. Our results demonstrate that the presence of exotic matter fields characterized by the parameters $\alpha$ and $\lambda$ influences the quasinormal mode spectrum. In particular, increasing the quintessence parameter $\alpha$ induces a reduction in the oscillation frequency, reflecting a gravitational shielding effect, while subtle variations in the damping rate alter the relaxation dynamics. The string cloud parameter $\lambda$ modulates the structure of the effective potential, and consequently affects both the stability of perturbations and the characteristic scales of the system.

A systematic verification of Hod's conjecture revealed that this relaxation bound is satisfied over a substantial region of the parameter space. However, for specific critical combinations of $\alpha$ and $\lambda$, deviations from the bound emerge, signaling transitions in the thermodynamic behavior of the black hole in which the damping rate exceeds the Hawking temperature-controlled limit. These regimes point to a breakdown of the standard semiclassical description and merit further investigation. Analysis of the optical properties revealed strong correlations between the spacetime geometry of and observable astrophysical signatures. The expansion of the shadow radius with the quintessence parameter $\alpha$ can be interpreted as a manifestation of the gravitational potential deepened by quintessence, while the presence of the string cloud enhances the photon capture cross section, indicating that it contributes effectively to the gravitational mass of the black hole. The energy emission rate, meanwhile, exhibits characteristic spectral changes under the combined influence of these exotic fields, including a reduction in intensity and a shift in the emission peak, which may lead to observable implications.

The central result of this work is the incorporation of the Swampland distance conjecture into the black hole framework, establishing a direct connection between scalar field excursions and black hole thermodynamics. The formulation of a modified Hod bound, which combines semiclassical relaxation constraints with quantum gravity consistency conditions, delimits a viable region of consistency in the parameter space in which both the semiclassical and effective descriptions remain valid. This synthesis opens up promising prospects for the study of the interfaces between general relativity, cosmology, and quantum gravity theories.

In conclusion, this work lays the groundwork for the exploration of charge black holes in enriched astrophysical environments, where the interaction between gravity, exotic fields, and fundamental constraints of quantum gravity shapes a rich and potentially testable phenomenology. Future extensions of cloud include the effects of rotation, nonlinear stability analysis, or the search for specific observational signatures in contemporary astrophysical data.

Finally, an intriguing open problem addresses the ultimate source of the relaxation bound itself. Although Hod's conjecture has been tested extensively in semiclassical black hole physics, its relation to quantum gravity consistency conditions remains unclear. As far as the Swampland program is concerned, it is natural to wonder if such relaxation bounds could be some sort of constraint on low–energy effective field theories that can admit a consistently ultraviolet completion.

In this context, violation of the Hod bound in regions of the parameter space can imply that the effective descriptions in those regions may be outside the set of quantum gravity compatible theories. An interesting future direction of research would be to investigate the relationship between the black hole relaxation bounds and the Swampland constraints.

\section*{Acknowledgments}
The work of S. Saoud, M. A. Rbah and R. Sammani  is funded by the National Center for Scientific and Technical
Research (CNRST) under the PhD-Associate Scholarship (PASS).
\bibliographystyle{unsrt}
\bibliography{Bib}
\end{document}